\documentclass[preprintnumbers,
amsmath,amssymb,floatfix,10pt,prd,onecolumn,
superscriptaddress,longbibliography,nofootinbib]{revtex4-2}
\usepackage{bm}
\usepackage{amsfonts}
\usepackage{latexsym}
\usepackage{amsmath}
\usepackage{textcomp}
\usepackage{float}
\usepackage{booktabs}
\usepackage{dcolumn}
\usepackage{dsfont}
\usepackage{ragged2e}
\usepackage{epsfig}
\usepackage[dvipsnames]{xcolor}
\usepackage{hyperref}
\hypersetup{           
	colorlinks=true,                
	breaklinks=true,                
	urlcolor= OrangeRed,                
	linkcolor= magenta,                
	bookmarksopen=false,
	filecolor=black,
	citecolor=red,
	linkbordercolor=blue}
\usepackage{graphicx}
\usepackage{overpic}
\usepackage{orcidlink}
\usepackage[T1]{fontenc}

\begin{document}
	\title{Dynamics of Late time cosmology in $f(Q,L_{m})$ Gravity with Constraints from DESI DR2 BAO Data}
	
   \author{Rajdeep Mazumdar \orcidlink{0009-0003-7732-875X}}
	\email[Corresponding author: ]{rajdeepmazumdar377@gmail.com}
	\affiliation{%
		Department of Physics, Dibrugarh University, Dibrugarh, Assam, India, 786004}

  \author{Kalyan Malakar\orcidlink{0009-0002-5134-1553}}%
\email{kalyanmalakar349@gmail.com}
\affiliation{Department of Physics, Dibrugarh University, Dibrugarh, Assam, India, 786004}
\affiliation{Department of Physics, Silapathar College, Dhemaji, Assam, India, 787059}
	
	\author{Kalyan Bhuyan\orcidlink{0000-0002-8896-7691}}%
	\email{kalyanbhuyan@dibru.ac.in}
	\affiliation{%
		Department of Physics, Dibrugarh University, Dibrugarh, Assam, India, 786004}%
	\affiliation{Theoretical Physics Divison, Centre for Atmospheric Studies, Dibrugarh University, Dibrugarh, Assam, India 786004}

	
\begin{abstract}
We investigate late-time cosmology in the context of modified $f(Q,L_m)$ gravity, considering a non-linear model$
f(Q,L_m) = \alpha Q + \beta L_m^n + \lambda$
where, $\alpha$, $\beta$, $\lambda$, and $n$ are some free parameters. The modified Friedmann equations are derived for a barotropic cosmic fluid, and an analytical solution for the Hubble parameter $H(z)$ is obtained. Using the latest DESI DR2 BAO data, previous BAO compilations (P-BAO), and cosmic chronometer (CC) datasets, we constrain the model parameters through a Markov Chain Monte Carlo analysis. Our results show that the model successfully describes the observed late-time cosmic acceleration with slightly tighter constraints from the inclusion of DESI dataset. The present-day Hubble constant is determined as $H_0 \simeq 69.5\ \mathrm{km\ s^{-1}\ Mpc^{-1}}$, while the deceleration parameter confirms accelerated expansion with $q_0 \simeq -0.57$. The transition redshift, where the universe switches from deceleration to acceleration, occurs in the range $z_{\rm tr} \sim 0.56 - 0.77$. Similarly, a smooth and physically consistent transition from a matter-dominated decelerated period at high redshifts to an accelerated phase at late times is revealed by the evolution of $\omega_{eff}(z)$. 
While statefinder diagnostic shows the model favours a Chaplygin gas like nature for DESI and DESI+CC, whereas the model favours as quintessence dominated evolution for P-BAO+CC in the late time regime. Conclusively, all these results along with the study of the energy conditions and stability analysis showcases the given $f(Q,L_m)$ model offers a viable alternative to GR-based cosmology
\end{abstract}

	
	\maketitle

\section{Introduction}\label{s1}
The revelation that the Universe is currently experiencing a phase of late-time accelerated expansion stands as one of the most significant milestones in contemporary cosmology. A growing number of independent observational probes, such as high-precision measurements of the Cosmic Microwave Background Radiation (CMBR) \cite{Spergel2003}, Baryon Acoustic Oscillations (BAO), and extensive large-scale structure (LSS) surveys \cite{Tegmark2004, Cole2005, Eisenstein2005}, have subsequently confirmed this remarkable discovery, which was first suggested by observations of Type Ia Supernovae (SNe Ia) \cite{Riess1998, Perlmutter1999, Astier2006}. When taken as a whole, these results demonstrate that the Universe is not only expanding but it's expansion is accelerating over the time. Researchers in diverse studies have postulated the presence of a mysterious component called dark energy (DE), which is described as a kind of energy with enormous negative pressure that accounts for around 70\% of the cosmic energy content, in order to explain this accelerated expansion of the universe. The traditional $\Lambda$CDM model, which has been shown to be in great agreement with an extensive range of observational evidence spanning many cosmic epochs, is based on the simplest and most frequently accepted candidate for DE, the cosmological constant ($\Lambda$). However, the $\Lambda$CDM model has a number of conceptual flaws despite its extensive empirical robustness or resilience. Foremost among these is the ``cosmological constant problem'', which arises from the startling difference between the values of vacuum energy density that have been determined from observations and those that are predicted theoretically \cite{Weinberg1989, Martin2012}. Equally significant are the ``fine-tuning'' and ``cosmic coincidence problems'', which challenge our understanding of why the present-day dark energy density is so delicately adjusted and why its dominance coincides precisely with the current cosmological epoch. These persistent theoretical difficulties and shortcomings have inspired the development of alternative frameworks to explain cosmic acceleration without inserting an ad hoc $\Lambda$ as cosmological constant. Either through the introduction of dynamical dark energy models with evolving equations of state or via modifications to Einstein’s theory of gravity. A comprehensive review of these directions can be found in Refs.~\cite{za1}--\cite{za10}.\\
By extending or reformulating conventional general relativity, modified gravity theories aim to provide a more precise and comprehensive explanation of the dynamics of the universe on large-scale \cite{p13}. Modified gravity theories are seen to be fundamentally effective to solve significant cosmological issues that the normal GR framework is still unable to adequately explain,  such dark energy, dark matter, and the universe's late-time accelerated expansion. These theories are also seen to provide new insights into the fundamental dynamics driving cosmic evolution. Such modified theories of gravity can be obtained by generalizing the Einstein-Hilbert action through the introduction of new components that depend on geometric characteristics like curvature invariants, torsion, or non-metricity, which can enhance the structure of the theories and increases their ability for precise predictions. Basically the fundamental basis of spacetime curvature in GR is formulated through the Riemann tensor. One of the earliest and most widely studied modified gravity theory is the ``$f(R)$ gravity theory'', which substitutes a generic function $f(R)$ \cite{p14} for the Ricci scalar $R$ in the Einstein–Hilbert action. Simultaneously, other alternative geometric frameworks have also been introduced that reformulate the gravity through torsion or non-metricity instead of curvature. Most popular among these are the ``Teleparallel Equivalent of General Relativity'' (TEGR) \cite{p16}, which makes use of the torsion scalar $T$, and the ``Symmetric Teleparallel Equivalent of General Relativity'' (STEGR) \cite{p17,p18}, which which makes use of the non-metricity scalar $Q$. These frameworks have lead to the formulation of the $f(T)$ and $f(Q)$ gravity theories \cite{p19,p20,p21}, where the gravitational Lagrangian is generalized using some arbitrary functions of $T$ and $Q$, respectively. Further developments have been made by incorporating additional couplings and functional dependencies, that are seen to propose even richer theoretical structures. Notable examples include $f(R,T)$ \cite{p22}, $f(R,L_m)$ \cite{p23,p24}, $f(T,\tau)$ \cite{p25}, $f(T,\phi)$ \cite{p26}, $f(T,B)$ \cite{p28}, and $f(Q,T)$ gravity \cite{p29}. These formulations are seen to offer new ways for studying non-minimal interactions between geometry and matter, which may result in novel observable phenomena and gravitational dynamics. A particularly intriguing extension in this direction is the $f(Q,L_m)$ gravity theory \cite{pp45}, where the matter Lagrangian $L_m$ is directly coupled to the non-metricity scalar $Q$. Without the need for an exotic dark energy components (such as $\Lambda$), this non-minimal coupling creates an interaction between the geometric and material sectors of the universe, offering a novel explanation for the accelerated expansion of the universe and other late-time events. The flexibility of the functional form of $f(Q,L_m)$  also makes it possible to formulate modelsthat can adequately describe both early and late-time cosmological dynamics of the universe. Our understanding of the $f(Q,L_m)$ gravity framework and its cosmological implications has been greatly expanded by a number of recent studies. Such as a thorough discusiion of the overall theoretical framework of $f(Q,L_m)$ gravity, the derivation of the associated field equations, and a study of their cosmological applications was provided by Hazarika \textit{et al.}~\cite{pp45} in his recent stuides. K. Myrzakulov \textit{et al.}~\cite{pp46} performed an observational analysis of late-time cosmic acceleration under this framework using a range of cosmology datasets in order to constrain the model parameters. Y. Myrzakulov \textit{et al.}~\cite{pp47} found analytical solutions characterising the late-time dynamics of the Universe in the $f(Q,L_m)$ framework and evaluated how well their predictions matched observational data. Additionally, Y. Myrzakulov \textit{et al.}~\cite{pp48} also investigated the bulk viscosity effects in this framework, showing how such additional processes might affect the pace of cosmic expansion. More recently, Samaddar and Singh~\cite{pp49}  studied the vaibility of  baryogenesis within the $f(Q,L_m)$ framework, which highlighted its possible significance for early-Universe physics. Samaddar \& Surendra~\cite{pp33} have also examined the phase-space behaviour of Barrow Agegraphic Dark Energy (BADE) in $f(Q,L_m)$ gravity, demonstrating the theory’s ability to unify different epochs of cosmic evolution. Recent publications \cite{pp45,n1} have also highlighted the potential of $f(Q,L_m)$ gravity in resolving theoretical and empirical cosmological problems more extensively. These investigations suggested that the connection between non-metricity and matter within the framework of $f(Q,L_m)$ gravity may successfully mimic the late-time cosmic acceleration while remaining consistent with current observational constraints. Collectively, these studies highlight the theoretical richness and empirical adaptability of $f(Q,L_m)$ gravity, encouraging further investigation into its dynamical behaviour and effects on cosmology of the universe.\\
Motivated by these advances, the current study intends to investigate the late-time cosmological dynamics of $f(Q,L_m)$ gravity in more detail and compare its predictions with the most recent observational data from the Baryon Acoustic Oscillation (BAO) measurements made with the Dark Energy Spectroscopic Instrument (DESI) Data Release 2 (DR2). In particular, we consider specific and physically motivated functional forms of the arbitrary function $f(Q,L_m)$ and analyze their implications for cosmological evolution. By deriving the modified Friedmann equations and constructing the key dynamical quantities, we explore how the non-metricity–matter coupling influences the background expansion history. To test the observational viability of the proposed models, we employ the latest DESI DR2 BAO measurements, which provide highly precise constraints on the dynamics of the cosmic expansion. We try to provide new insights into the phenomenological potential of $f(Q,L_m)$ gravity as an alternative to the standard $\Lambda$CDM scenario by constraining the model parameters and evaluating its ability to explain the observed late-time cosmic acceleration using extensive statistical analysis and the most recent high precession datasets. The findings of this work thus aim to represent a significant step toward establishing $f(Q,L_m)$ gravity as a robust contender in modern cosmology and contribute to the broader pursuit of identifying theoretically consistent and observationally supported extensions of General Relativity.\\
This paper is organised as follows: Section \ref{s2} gives a brief mathematical formalism of $f(Q,L_m)$ gravity; Section \ref{s3} illustrates the cosmological framework for $f(Q,L_m)$ gravity in a flat FLRW Universe; and Section \ref{s4} discusses the cosmological $f(Q,L_m)$ model considered in the study. The methodology and observational data utilised in the study to constrain the model are covered in Sec. \ref{s5}. The constraints on the model and their cosmological implications are covered in Sec \ref{s6}. Lastly, a summary of the study is given in Sec. \ref{s6}.
\section{The $f(Q,L_m)$ Gravity Theory}\label{s2}
In this work, we consider an extended formulation of symmetric teleparallel gravity in which the gravitational action depends on both the non-metricity scalar \(Q\) and the matter Lagrangian
\(\mathcal{L}_m\). The gravitational action for it is given by:
\begin{equation}
S=\int f(Q,L_m)\sqrt{-g}\, d^4x,
\label{e1}
\end{equation}
where $f(Q,L_m)$ is some arbitrary function of the matter Lagrangian $L_m$ and the non-metricity scalar $Q$, and $g$ is the metric tensor's determinant. This framework generalizes both $f(Q)$ gravity and the standard GR formulation by allowing geometric-matter coupling at the level of the action.\\
The non-metricity scalar $Q$ is defined as\cite{q}:
\begin{equation}
Q=-g^{\mu\nu}\left(L^{\beta}{}_{\alpha\mu}L^{\alpha}{}_{\nu\beta}-L^{\beta}{}_{\alpha\beta}L^{\alpha}{}_{\mu\nu}\right),
\end{equation}
where $L^{\beta}{}_{\alpha\gamma}$ is the disformation tensor,
\begin{equation}
L^{\beta}{}_{\alpha\gamma}=\frac{1}{2} g^{\beta\eta}
\left( Q_{\gamma\alpha\eta}+Q_{\alpha\eta\gamma}-Q_{\eta\alpha\gamma} \right),
\end{equation}
and the nonmetricity tensor is defined as:
\begin{equation}
Q_{\gamma\mu\nu}=-\nabla_{\gamma}g_{\mu\nu}
=-\partial_{\gamma}g_{\mu\nu}
+g_{\nu\sigma}\, \tilde{\Gamma}^{\sigma}{}_{\mu\gamma}
+g_{\sigma\mu}\, \tilde{\Gamma}^{\sigma}{}_{\nu\gamma},
\end{equation}
where in symmetric teleparallel geometry, the link is represented by $\tilde{\Gamma}^{\sigma}{}_{\mu\nu}$. The nonmetricity tensor's traces are provided by:
\begin{equation}
Q_{\beta}=g^{\mu\nu}Q_{\beta\mu\nu}, 
\qquad 
\tilde{Q}_{\beta}=g^{\mu\nu}Q_{\mu\beta\nu}.
\end{equation}
We introduce the superpotential tensor, sometimes referred to as the non-metricity conjugate, to make the field equations easier:
\begin{widetext}
\begin{equation}
P^{\beta}{}_{\mu\nu} \equiv \frac{1}{4}
\left[
- Q^{\beta}{}_{\mu\nu}
+ 2 Q^{\beta}{}_{(\mu\nu)}
+ Q_{\beta} g_{\mu\nu}
- \tilde{Q}_{\beta} g_{\mu\nu}
- \delta^{\beta}_{(\mu} Q_{\nu)}
\right]
= -\frac{1}{2} L^{\beta}{}_{\mu\nu}
+ \frac{1}{4} (Q_{\beta} - \tilde{Q}_{\beta}) g_{\mu\nu}
- \frac{1}{4} \delta^{\beta}_{(\mu} Q_{\nu)}.
\end{equation}
\end{widetext}
In teleparallel gravity, this tensor has an operation similar to that of the contortion tensor. The non-metricity scalar can therefore be expressed as in terms of this conjugate tensor\cite{q}:
\begin{widetext}
\begin{equation}
Q = - Q_{\beta\mu\nu} P^{\beta\mu\nu}
= -\frac{1}{4}
\left(
- Q_{\beta\nu\rho} Q^{\beta\nu\rho}
+ 2 Q_{\beta\nu\rho} Q^{\rho\beta\nu}
- 2 Q_{\rho} \tilde{Q}^{\rho}
+ Q_{\rho} Q^{\rho}
\right).
\end{equation}
\end{widetext}
By varying the action in Eq. (\ref{e1}) with respect to the metric, one can obtain the gravitational field equations as:,
\begin{widetext}
\begin{equation}
2\sqrt{-g}\,\nabla_{\alpha}\!\left(f_{Q}\sqrt{-g}\,P^{\alpha}{}_{\mu\nu}\right)
+ f_{Q}\left(P_{\mu\alpha\beta} Q^{\alpha\beta}{}_{\nu}
- 2 Q_{\alpha\beta\mu} P^{\alpha\beta}{}_{\nu}\right)
+ \frac{1}{2} f\, g_{\mu\nu}
= \frac{1}{2} f_{L_m} \left(g_{\mu\nu} L_{m} - T_{\mu\nu}\right),
\label{e8}
\end{equation}
\end{widetext}
where $f_{Q} = \frac{\partial f}{\partial Q}$ and $f_{L_m} = \frac{\partial f}{\partial L_m}$. One recovers the usual \(f(Q)\) field equations when \(f(Q,L_m)=f(Q)+2L_m\) \cite{q}. The matter energy-momentum tensor is defined as:
\begin{equation}
T_{\mu\nu}
= g_{\mu\nu}L_m - 2\frac{\partial L_m}{\partial g^{\mu\nu}}.
\end{equation}
Again, variation of the action with respect to the connection leads to an extra equation:
\begin{equation}
\nabla_{\mu}\nabla_{\nu}
\left(4\sqrt{-g}\, f_Q P^{\mu\nu}{}_{\alpha}
+ H^{\mu\nu}{}_{\alpha}\right)=0,
\end{equation}
where the hypermomentum density is given by
\begin{equation}
H^{\mu\nu}{}_{\alpha}
= \sqrt{-g}\, f_{L_m} \frac{\partial L_m}{\partial Y^{\alpha}{}_{\mu\nu}}.
\end{equation}
A distinctive aspect of \(f(Q,\mathcal{L}_m)\) gravity is the generic non-conservation of the energy–momentum tensor. Taking the covariant divergence of the field equations \ref{e8} leads to:
\begin{widetext}
\begin{equation}
\nabla_{\mu} T^{\mu}{}_{\nu}
=\frac{1}{f_{\mathcal{L}_m}}
\left[
2\nabla_\alpha \nabla_\mu H^{\alpha\mu}{}_{\nu}
+\nabla_\mu A^{\mu}{}_{\nu}
-\nabla_\mu \left(\frac{1}{\sqrt{-g}}
\nabla_\alpha H^{\alpha\mu}{}_{\nu}\right)
\right]
\equiv B_\nu \neq 0,
\end{equation}
\end{widetext}
indicating that an exchange of energy and momentum occurs between the matter sector
and the geometric sector through the non-minimal coupling. The explicit form of \(B_\nu\)
depends on the choice of \(f(Q,\mathcal{L}_m)\), the matter Lagrangian, and the dynamical variables involved. For $B_{\nu}=0$, standard conservation is recovered.
\section{Modified Friedmann Equations in $f(Q,L_m)$ Gravity}\label{s3}
In order to study the cosmic dynamics of $f(Q,L_m)$ gravity, we take into consideration an isotropic and spatially homogeneous universe that is characterised by the flat Friedmann–Lemaître–Robertson–Walker (FLRW) metric \cite{ref50}:
\begin{equation}
    ds^{2} = -dt^{2} + a^{2}(t)\left(dx^{2} + dy^{2} + dz^{2}\right),
    \label{metric}
\end{equation}
where the cosmic scale factor is denoted by $a(t)$. The Hubble parameter, $H = \dot{a}/a$, measures the universe's rate of expansion. For the metric as given above by Eq. (\ref{metric}), the non-metricity scalar becomes:
\begin{equation}
    Q = 6H^{2}.
\end{equation}
Here, we assume the matter sector to be a perfect fluid with energy–momentum tensor given as:
\begin{equation}
    T_{\mu\nu} = (\rho + p)u_{\mu}u_{\nu} + p\, g_{\mu\nu},
\end{equation}
where $u^{\mu}$ is the fluid's four-velocity, $p$ is the pressure, and $\rho$ is the energy density. The modified Friedmann equations in $f(Q,L_m)$ gravity that arise from the variation of the action using the metric and matter definitions mentioned above are \cite{pp45,pp46,pp47,pp48}:
\begin{equation}
    3H^{2} = \frac{1}{4 f_Q}\left[f - f_{L_m}(\rho + L_m)\right], 
    \label{fried1}
\end{equation}
\begin{equation}
    \dot{H} + 3H^{2} + \frac{\dot{f}_Q}{f_Q}H = \frac{1}{4 f_Q}\left[f + f_{L_m}(p - L_m)\right].
    \label{fried2}
\end{equation}
The generalized energy balance equation takes the form
\begin{equation}
    \dot{\rho} + 3H(\rho + p) = B_{\mu}u^{\mu},
\end{equation}
where the non-zero source term $B_{\mu}u^{\mu}$ indicates that the usual conservation of energy–momentum is violated. The condition $B_{\mu}u^{\mu}=0$ restores standard conservation; otherwise, it encodes a coupling between matter and geometry.  
Combining Eq. \eqref{fried1} and Eq. \eqref{fried2}, one obtains
\begin{equation}
    2\dot{H} + 3H^{2} = 
    \frac{1}{4 f_Q}
    \left[
        f + f_{L_m}(\rho + 2p - L_m)
    \right]
    - 2\frac{\dot{f}_Q}{f_Q}H.
\end{equation}
For further convenience, these modified Friedmann equations may be expressed in terms of effective thermodynamic quantities:
\begin{equation}
    3H^{2} = \rho_{\text{eff}}, \qquad
    2\dot{H} + 3H^{2} = -p_{\text{eff}}.
\end{equation}
The corresponding effective energy density and pressure are given by
\begin{equation}
    \rho_{\text{eff}} = \frac{1}{4 f_Q}\left[f - f_{L_m}(\rho + L_m)\right],
\end{equation}
\begin{equation}
    p_{\text{eff}} = 
    2\frac{\dot{f}_Q}{f_Q}H 
    - \frac{1}{4 f_Q}
      \left[f + f_{L_m}(\rho + 2p - L_m)\right].
\end{equation}
Finally, the effective continuity equation takes the standard form:
\begin{equation}
    \dot{\rho}_{\text{eff}} 
    + 3H\left(\rho_{\text{eff}} + p_{\text{eff}}\right) = 0.
\end{equation}
These modified cosmological equations form the basis for analyzing the dynamical behavior of the Universe in $f(Q,L_m)$ gravity. Their implications for cosmic evolution and observational constraints will be explored in the following sections.
\section{Cosmological Model}\label{s4}
Within the framework of $f(Q,L_m)$ gravity, we explore an extended gravity model in this study that is specified as:
\begin{equation}
f(Q,\mathcal{L}_m) = \alpha Q + \beta \mathcal{L}_m^{\,n} + \lambda ,
\end{equation}
where the independent model parameters are $\alpha$, $\beta$, $\lambda$, and $n$. Through a power-law dependency, this functional form creates a non-minimal coupling between the matter Lagrangian $L_m$ and the non-metricity scalar $Q$. Richer cosmic dynamics can be made possible by this choice, which generalises the minimal coupling scenario. The degree of geometry-matter coupling is determined by the parameter $n$. The model simplifies to a linear matter coupling for $n=1$, but effective interactions that can potentially alter the history of cosmic expansion are induced for $n \neq 1$. In different modified gravity theories, such $f(R,\mathcal{L}_m)$ and $f(R,\mathcal{L}_m,T)$, identical non-minimal matter couplings have been thoroughly investigated. These couplings are known to result in late-time acceleration and deviations from standard cosmic dynamics and their implications.\\
We assume a barotropic cosmic fluid throughout this study, which is represented by the equation of state:
\begin{equation}
p = \omega \rho ,
\end{equation}
where $\omega$ is some constant equation-of-state parameter. Accordingly, here the matter Lagrangian is taken as:
\begin{equation}
\mathcal{L}_m = \rho .
\end{equation}
It is worth noting that the $f(Q,\mathcal{L}_m)$ model used in this study shares some comparable similarities with those investigated by K. Myrzakulov et al. \cite{pp46} and Y. Myrzakulov et al. \cite{pp47, n1}. However, our present formulation extends those previous studies in some key aspects. First, rather than limiting the analysis to a pressureless matter sector ($\omega=0$), we consider a general barotropic cosmic fluid ($p=\omega\rho$) allowing scope for a broader exploration of cosmic epochs. Second, we parametrized the coupling strength of the non-metricity scalar $Q$ by a free parameter $\alpha$, hence allowing controlled deviations from the canonical STEGR normalization. Third, the addition of a constant term $\lambda$ in the function may effectively acts as a dynamical vacuum contribution, which can modulate or influence the late-time cosmic acceleration. When combined with new high-precision observational datasets, these modifications may offer more theoretical flexibility, resulting in a richer cosmic dynamics and enabling a broader spectrum of expansion histories beyond those examined in previous $f(Q,\mathcal{L}_m)$ frameworks.\\
Under these assumptions, the modified Friedmann equations corresponding to Eq. (\ref{fried1}) and Eq. (\ref{fried2}) take the form:
\begin{equation}
3H^2 = \frac{\lambda +6 \alpha  H^2-2 \beta  n \rho ^n+\beta  \rho ^n}{4 \alpha },
\label{firede1}
\end{equation}

\begin{equation}
3H^2 + 2\dot{H} = \frac{\lambda +6 \alpha  H^2+2 \beta  n \omega  \rho ^n+\beta  \rho ^n}{4 \alpha }.
\label{firede2}
\end{equation}
It is essential to point out that the aforementioned equations reduce to the standard Friedmann equations of General Relativity when $\alpha = -1$, $\beta = 2$, $\lambda = 0$, and $n = 1$. Using Eq. (\ref{firede1}) we obtain:
\begin{equation}
\rho^{n} = \frac{\lambda -6 \alpha  H^2}{\beta  (2 n-1)}
\label{firede3}
\end{equation} 
Combining Eqs. (\ref{firede1}), (\ref{firede2}) and (\ref{firede3}), we obtain the evolution equation for the Hubble parameter as:
\begin{equation}
\dot{H} = \frac{n (\omega +1) \left(6 \alpha  H^2-\lambda \right)}{4 \alpha  (1-2 n)},
\label{Hdot}
\end{equation}
where $\omega = p/\rho$ as mentioned is just the equation-of state parameter. Now, to express the cosmological dynamics in terms of redshift $z$, we use the relation:
\begin{equation}
1+z = \frac{a_0}{a(t)}, \qquad a_0 = 1,
\end{equation}
and the identity
\begin{equation}
\frac{d}{dt} = -H(1+z)\frac{d}{dz}.
\end{equation}
Applying this transformation to Eq.~\eqref{Hdot}, we obtain
\begin{equation}
-H(1+z)\frac{dH}{dz}
=
\frac{n (\omega +1) \left(6 \alpha  H^2-\lambda \right)}{4 \alpha  (1-2 n)} .
\label{Hz_diff}
\end{equation}
Solving the above differential equation, the Hubble parameter as a function of redshift is obtained as:
\begin{widetext}
\begin{equation}
H^2(z) =
H_0^2 (1+z)^{\frac{3n(1+\omega)}{2n-1}}
+ \frac{\lambda}{6\alpha}\left(1-(1+z)^{\frac{3n(1+\omega)}{2n-1}}\right)
,
\label{Hz1}
\end{equation}
\end{widetext}
where $H_0 = H(z=0)$ is the value of the Hubble parameter in the present time. Eq. (\ref{Hz1}) shows that the Hubble parameter have dependence on the redshift $z$, the model parameter $n$, the present Hubble constant $H_0$, and the parameters $\lambda$ and $\alpha$. Notably, the first term is independent of $\alpha$, and $\alpha$ appears only in the denominator of the second term as a scaling factor. This indicates that while $\alpha$ does not directly affect the redshift evolution encoded in the first term, it influences the magnitude of the additive contribution from the second term, thereby affecting the overall cosmic expansion history indirectly through $\lambda / \alpha$. For simplification of the study we replace $\frac{\lambda}{6\alpha}$ with $\gamma$ such that $\gamma = \frac{\lambda}{6\alpha}$, which give us the Hubble parameter $H(z)$ as:
\begin{equation}
H^2(z) =
H_0^2 (1+z)^{\frac{3n(1+\omega)}{2n-1}}
+ \gamma\left(1-(1+z)^{\frac{3n(1+\omega)}{2n-1}}\right)
,
\label{Hz}
\end{equation}\\
Additionally, after obtaining $H(z)$, we may evaluate the deceleration parameter, which is as follows:
\begin{equation}
    q(z) = -1 - \frac{\dot{H}}{H^2} = -1 + (1+z) \frac{1}{H(z)} \frac{dH}{dz},
    \label{eq1}
\end{equation}
It shows how rapidly the universe is expanding or contracting. Accelerated expansion is indicated by a negative value of $q(z)$. Additionally, we may get the effective EoS, which is as follows:
\begin{equation}
\omega_{eff}(z) = \frac{p_{eff}}{\rho_{eff}}=-1 + \frac{2(1+z)}{3H(z)}\frac{dH}{dz},
 \label{ew1}
\end{equation}
This offers an effective fluid description of the Universe's dynamics that incorporates both matter and dark energy effects, with $p_{eff}$ and $\rho_{eff}$ standing for the effective pressure and energy density. These values enable us to evaluate if the the phantom boundary ($w_{\text{eff}} < -1$) is crossed by the effective equation of state and whether the model predicts a change from deceleration to acceleration. Additionally, we may use the statefinder diagnostic pair $(r,s)$ alongside with the $(r,q)$ pair \cite{ref97n} to analyse the expansion history of the Universe and the nature of dark energy at late periods.  Using the scale factor $a(t)$ and its higher-order derivatives the statefinder parameters are defined as:
\begin{equation}
r = \frac{\dddot{a}}{aH^3},
\end{equation}
\begin{equation}
s = \frac{r - 1}{3\left(q - \frac{1}{2}\right)},
\label{es}
\end{equation}
Explicitly using $q$, the statefinder parameter $r$ can be written as:
\begin{equation}
r = 2q^2 + q - (1+z)\frac{dq}{dz}.
\label{er}
\end{equation}
In subsequent sections , we evaluate the constraints on the model parameters using some observational datasets. Based on these constraints we investigate the cosmological implications expressed by the aforementioned cosmological parameters , together with a thorough examination of energy conditions and stability, offering important insights into the behaviour of the model. 
\section{Observational Data and Statistical Methodology for Constraints}\label{s5}
For this study using Baryon Acoustic Oscillation (BAO) measurements and cosmic chronometer (CC) data, we constrain the free parameters of the given cosmological model. The BAO information is taken from two independent compilations: (i) the recent DESI DR2 BAO dataset \cite{ref98}, and (ii) a previous BAO compilation (P-BAO) consisting of measurements from SDSS, WiggleZ, DES, and related surveys \cite{d113}--\cite{d79}. The CC data provide independent measurements of the Hubble parameter \(H(z)\) from the relative ages of passively-evolving galaxies \cite{x38} and serve as a complementary probe of the expansion history. To examine the constraining power and internal consistency of the datasets, we perform three separate analyses throughout this work:
\begin{itemize}
\item DESI-only,
\item DESI + CC,
\item P-BAO + CC.
\end{itemize}
The BAO observables considered in this analysis include the Hubble parameter \(H(z)\), the Hubble distance \(D_H(z)\), and the transverse comoving distance \(D_M(z)\). In particular, the BAO distances are reported in units of the sound horizon at the drag epoch, \(r_d\). Here, the Hubble distance is defined as:
\begin{equation}
D_H(z) = \frac{c}{H(z)},
\end{equation}
where \(c\) denotes the speed of light in  and $H(z)$ is the Hubble parameter. The transverse comoving distance is given by:
\begin{equation}
D_M(z) = c \int_0^z \frac{dz'}{H(z')}.
\end{equation}
Both the DESI DR2 BAO dataset and the previous BAO compilation (P-BAO), along with CC Dataset provide measurements of these cosmological observables or quantities, allowing us to probe the expansion history of the Universe across a wide redshift range. In the present work, we use the normalized quantities \(D_H(z)/r_d\) and \(D_M(z)/r_d\), together with direct measurements of \(H(z)\), to construct the likelihood functions for the DESI-only, DESI+CC, and P-BAO+CC analyses. The corresponding observational data points and uncertainties are summarized in Tables~\ref{tab1a}, \ref{tab1b}, and \ref{tab1c}. Sound waves travelling in the strongly coupled photon--baryon plasma prior to the drag epoch give rise to Baryon Acoustic Oscillations (BAO). The sound horizon during the drag epoch ($r_d$) is a distinctive comoving length scale left by these oscillations after decoupling. It is imprinted on both the large-scale distribution of matter and the Cosmic Microwave Background (CMB) anisotropies. BAO observations offer a reliable standard measurement for investigating the cosmic expansion history across a broad redshift range because of its geometric character and low susceptibility to astrophysical systematics. Although other cosmic probes, including Type Ia supernovae etc. can also be utilized for obtaining the constraints on the model parameters, but we limit our attention to the aforementioned datasets only for simplicity of the task. Through the datasets as discussed above the best fitting parameter values for the given model are obtained using the standard chi-squared minimization method. With respect to which different $\chi^2$ functions used in this work for statical analysis are defined as following:\\
\begin{table}[h!]
\centering
\begin{tabular}{|ccc|c||ccc|c|}
\hline
 \multicolumn{4}{|c||}{\textbf{P-BAO}} & \multicolumn{4}{c|}{\textbf{CC}} \\
\hline
$z$ & $H(z)$ & $\sigma_H$ & Ref & $z$ & $H(z)$ & $\sigma_H$ & Ref \\
\hline
0.24 & 79.69 & 2.99 & \cite{d113} & 0.07 & 69.00 & 19.60 & \cite{cc54} \\
 0.30 & 81.70 & 6.22 & \cite{d114} & 0.09 & 69.00 & 12.00 & \cite{cc55} \\
 0.31 & 78.17 & 6.74 & \cite{d115} & 0.12 & 68.60 & 26.20 & \cite{cc54} \\
0.34 & 83.17 & 6.74 & \cite{d113} & 0.17 & 83.00 & 8.00 & \cite{cc55} \\
 0.35 & 82.70 & 8.40 & \cite{d116} & 0.179 & 75.00 & 4.00 & \cite{cc56} \\
0.36 & 79.93 & 3.39 & \cite{d115} & 0.199 & 75.00 & 5.00 & \cite{cc56} \\
0.38 & 81.50 & 1.90 & \cite{d5} & 0.20 & 72.90 & 29.60 & \cite{cc54} \\
 0.40 & 82.04 & 2.03 & \cite{d115} & 0.27 & 77.00 & 14.00 & \cite{cc55} \\
 0.43 & 86.45 & 3.68 & \cite{d113} & 0.28 & 88.80 & 36.60 & \cite{cc54} \\
 0.44 & 82.60 & 7.80 & \cite{d74} & 0.352 & 83.00 & 14.00 & \cite{cc56} \\
0.44 & 84.81 & 1.83 & \cite{d115} & 0.3802 & 83.00 & 13.50 & \cite{cc58} \\
0.48 & 87.79 & 2.03 & \cite{d115} & 0.4 & 95.00 & 17.00 & \cite{cc55} \\
0.56 & 93.33 & 2.32 & \cite{d115} & 0.4004 & 77.00 & 10.20 & \cite{cc58} \\
0.57 & 87.60 & 7.80 & \cite{d10} & 0.4247 & 87.10 & 11.20 & \cite{cc58} \\
 0.57 & 96.80 & 3.40 & \cite{d117} & 0.4497 & 92.80 & 12.90 & \cite{cc58} \\
0.59 & 98.48 & 3.19 & \cite{d115} & 0.47 & 89.00 & 50.00 & \cite{cc59} \\
0.60 & 87.90 & 6.10 & \cite{d74} & 0.4783 & 80.90 & 9.00 & \cite{cc58} \\
 0.61 & 97.30 & 2.10 & \cite{d5} & 0.48 & 97.00 & 62.00 & \cite{cc59} \\
 0.64 & 98.82 & 2.99 & \cite{d115} & 0.593 & 104.00 & 13.00 & \cite{cc56} \\
0.978 & 113.72 & 14.63 & \cite{d118} & 0.68 & 92.00 & 8.00 & \cite{cc56} \\
1.23 & 131.44 & 12.42 & \cite{d118} & 0.781 & 105.00 & 12.00 & \cite{cc56} \\
 1.48 & 153.81 & 6.39 & \cite{d79} & 0.875 & 125.00 & 17.00 & \cite{cc56} \\
1.526 & 148.11 & 12.71 & \cite{d118} & 0.88 & 90.00 & 40.00 & \cite{cc59} \\
 1.944 & 172.63 & 14.79 & \cite{d118} & 0.9 & 117.00 & 23.00 & \cite{cc55} \\
 2.30 & 224.00 & 8.00 & \cite{d119} & 1.037 & 154.00 & 20.00 & \cite{cc56} \\
2.36 & 226.00 & 8.00 & \cite{d120} & 1.3 & 168.00 & 17.00 & \cite{cc55} \\
2.40 & 227.80 & 5.61 & \cite{d121} & 1.363 & 160.00 & 33.60 & \cite{cc60} \\
 &  &  &  & 1.43 & 177.00 & 18.00 & \cite{cc55} \\
 &  &  &  & 1.53 & 140.00 & 14.00 & \cite{cc55} \\
 &  &  &  & 1.75 & 202.00 & 40.00 & \cite{cc55} \\
 &  &  &  & 1.965 & 186.50 & 50.00 & \cite{cc60} \\
\hline
\end{tabular}
\caption{Hubble parameter $H(z)$ and its uncertainty at redshift $z$ from the Cosmic Chronometer (CC), DESI, and P-BAO datasets, measured in units of $km s^{-1} Mpc^{-1}$.}
\label{tab1a}
\end{table}

\begin{table}[h!]
\centering
\renewcommand{\arraystretch}{1.5}
\begin{tabular}{c|c|c|c|c}
\hline\hline
Tracer & $z_{\rm eff}$ & $D_M(z)/r_d$ & $D_H(z)/r_d$ & $r_{M,H}$ \\
\hline
LRG1 & 0.510 & 13.588 $\pm$ 0.167 & 21.863 $\pm$ 0.425 & $-0.459$ \\
LRG2 & 0.706 & 17.351 $\pm$ 0.177 & 19.455 $\pm$ 0.330 & $-0.404$ \\
LRG3+ELG1 & 0.934 & 21.576 $\pm$ 0.152 & 17.641 $\pm$ 0.193 & $-0.416$ \\
ELG2 & 1.321 & 27.601 $\pm$ 0.318 & 14.176 $\pm$ 0.221 & $-0.434$ \\
QSO & 1.484 & 30.512 $\pm$ 0.760 & 12.817 $\pm$ 0.516 & $-0.500$ \\
Ly$\alpha$ & 2.330 & 38.988 $\pm$ 0.531 & 8.632  $\pm$ 0.101 & $-0.431$ \\
\hline\hline
\end{tabular}
\caption{DESI dataset used in this analysis, which includes the transverse comoving distance
$D_M(z)/r_d$ and the Hubble distance $D_H(z)/r_d$, together with their respective uncertainties ans correlation coefficient $r_{M,H}$ for each effective redshift $z_{eff}$ \cite{ref98}.}
\label{tab1b}
\end{table}

\begin{table}[h!]
\centering
\renewcommand{\arraystretch}{1.5}
\begin{tabular}{c|c|c|c}
\hline\hline
$z_{\rm eff}$ & Observable & Value & Reference \\
\hline
0.38 & $D_M/r_d$ & $10.272 \pm 0.135 \pm 0.074$ & \cite{w59} \\
0.51 & $D_M/r_d$ & $13.378 \pm 0.156 \pm 0.095$ & \cite{w59} \\
0.61 & $D_M/r_d$ & $15.449 \pm 0.189 \pm 0.108$ & \cite{w59} \\
0.698 & $D_M/r_d$ & $17.65 \pm 0.30$ & \cite{w56} \\
1.48 & $D_M/r_d$ & $30.21 \pm 0.79$ & \cite{w57} \\
2.30 & $D_M/r_d$ & $37.77 \pm 2.13$ & \cite{w55} \\
2.40 & $D_M/r_d$ & $36.60 \pm 1.20$ & \cite{w60}\\
\hline
0.698 & $D_H/r_d$ & $19.77 \pm 0.47$ & \cite{w56} \\
1.48 & $D_H/r_d$ & $13.23 \pm 0.47$ & \cite{w57} \\
2.30 & $D_H/r_d$ & $9.07 \pm 0.31$ & \cite{w55} \\
2.40 & $D_H/r_d$ & $8.94 \pm 0.22$ & \cite{w60} \\
\hline\hline
\end{tabular}
\caption{The transverse comoving distance
$D_M(z)/r_d$ and the Hubble distance $D_H(z)/r_d$, together with their respective uncertainties for each effective redshift $z_{eff}$ from Previous BAO (P-BAO) dataset used in this analysis.}
\label{tab1c}
\end{table}
\subsection{CC Dataset}
Hubble parameter $H(z)$ measurement are provided by the CC data with uncorrelated Gaussian uncertainties.The chi-square for the CC dataset is therefore written as:
\begin{equation}
\chi^2_{\rm CC}(\theta) = \sum_j \frac{\left[H^{\rm obs}(z_j)-H^{\rm th}(z_j,\theta)\right]^2}{\sigma_{H,j}^2}.
\end{equation}
where the observed Hubble parameter and its uncertainty at redshift \(z_j\) are denoted by \(H^{\rm obs}(z_j)\) and \(\sigma_{H,j}\). While \(H^{\rm th}(z_j,\theta)\) denotes the theoretically predicted value of the Hubble parameter at the redshift $z_j$ and $\theta$ denotes the model parameters on which give theoretical value depends.
\subsection{DESI Dataset}
For the DESI DR2 dataset, BAO constraints are to be provided through the correlated observables \(D_M(z)/r_d\) and \(D_H(z)/r_d\), along with their corresponding uncertainties and correlation coefficients. Consequently, the statistical analysis requires a full covariance treatment. For each redshift bin \(z_i\), we construct the covariance matrix
\begin{equation}
\mathbf{C}_i =
\begin{pmatrix}
\sigma^2_{D_M} & r_{HM}\sigma_{D_M}\sigma_{D_H} \\
r_{HM}\sigma_{D_M}\sigma_{D_H} & \sigma^2_{D_H}
\end{pmatrix},
\end{equation}
where \(\sigma_{D_M}\) and \(\sigma_{D_H}\) denote the observational uncertainties associated to
\(D_M(z)/r_d\) and \(D_H(z)/r_d\), respectively, and \(r_{HM}\) is their correlation coefficient between the two observables. The corresponding chi-square estimator for the DESI dataset is then defined as
\begin{equation}
\chi^2_{\rm DESI}(\theta) = \sum_i \Delta_i^{\rm T} \mathbf{C}_i^{-1} \Delta_i,
\end{equation}
with
\begin{equation}
\Delta_i =
\begin{pmatrix}
D_M^{\rm obs}(z_i) - D_M^{\rm th}(z_i,\theta) \\
D_H^{\rm obs}(z_i) - D_H^{\rm th}(z_i,\theta)
\end{pmatrix}.
\end{equation}
\subsection{P-BAO Dataset}
The previous BAO compilation (P-BAO) provides measurements of the Hubble parameter \(H(z)\), the transverse comoving distance \(D_M(z)\), and the Hubble distance \(D_H(z)\), reported in units of the sound horizon \(r_d\). Unlike the DESI DR2 dataset, this compilation does not provide correlation coefficients between the radial and transverse BAO observables. Therefore, we construct the likelihood assuming diagonal covariance. The chi-square contribution from the P-BAO dataset is defined as:
\begin{widetext}
\begin{equation}
\chi^2_{\rm P\mbox{-}BAO}(\theta) =
\sum_i \frac{\left[D_M^{\rm obs}(z_i)-D_M^{\rm th}(z_i,\theta)\right]^2}{\sigma_{D_M,i}^2}
+
\sum_k \frac{\left[D_H^{\rm obs}(z_k)-D_H^{\rm th}(z_k,\theta)\right]^2}{\sigma_{D_H,k}^2}
+
\sum_j \frac{\left[H^{\rm obs}(z_j)-H^{\rm th}(z_j,\theta)\right]^2}{\sigma_{H,j}^2},
\end{equation}
\end{widetext}
where \(\sigma_{D_M,i}\), \(\sigma_{D_H,k}\), and \(\sigma_{H,j}\) denote the corresponding observational uncertainties.\\
Based on the above definitions, the total chi-square for each analysis is given by:
\begin{equation}
\chi^2_{\rm tot} =
\begin{cases}
\chi^2_{\rm DESI}, & \text{DESI}, \\
\chi^2_{\rm DESI} + \chi^2_{\rm CC}, & \text{DESI+CC}, \\
\chi^2_{\rm P\mbox{-}BAO} + \chi^2_{\rm CC}, & \text{P-BAO+CC},
\end{cases}
\end{equation}
The posterior distributions of $n$, $\gamma$, $\Omega$, and $H_0$ are then found by exploring the parameter space using a Markov Chain Monte Carlo (MCMC) technique. We use the following priors on the model parameters for the MCMC analysis: $H_0 \in [65, 75]$, $\Omega_{m0} \in [-0.5, 0]$, $n \in [0.5, 1.5]$and $\gamma \in [3400, 3500]$. The parameter values $(H_0, \Omega_{m0}, \gamma) = (70, -0.2 , 0.8, 3450)$ were used to initialise the MCMC chains. Using the three $\chi^2$ functions as previously described, we do the MCMC analysis in three steps.\\
We use a variety of statistical markers to evaluate the strength of the fits and compare the suggested model with the standard $\Lambda$CDM cosmology. For example, the Akaike Information Criterion (AIC), the Bayesian Information Criterion (BIC), the coefficient of determination ($R^2$), and the minimum chi-squared value ($\chi^2_{\min}$) (see Refs.~\cite{refS66}, \cite{refS68}, \cite{refS64} for details).One can define the Akaike Information Criterion (AIC) and the Bayesian Information Criterion (BIC) as:
\begin{equation}
\mathrm{AIC} = \chi^2_{\rm min} + 2k,
\qquad
\mathrm{BIC} = \chi^2_{\rm min} + k \ln N,
\end{equation}
where \(k\) denotes the number of free parameters and \(N\) denotes the total number of data points used in the analysis. A statistically preferred model is indicated by lower AIC and BIC values, with BIC offering a more severe penalty for model complexity. A model with lower AIC, BIC, and $\chi^2_{\min}$ values and a higher $R^2$ value is statistically preferable due to the displays of a better balance between complexity and quality of fit. Here, $R^2$ and $\chi^2_{\min}$, evaluate fit quality without taking the number of free parameters into consideration. On the other hand, models with more complexity are penalised by both AIC and BIC; however, BIC imposes a more severe penalty, particularly when dealing with bigger datasets. We calculate the differences in AIC and BIC as follows to assess the relative performance of models: $\Delta X = \Delta \text{AIC} = \mathrm{AIC}_{\Lambda\mathrm{CDM}}-\mathrm{AIC}_{\text{model}}$ or $ \Delta X = \Delta \text{BIC} = \mathrm{BIC}_{\Lambda\mathrm{CDM}}-\mathrm{BIC}_{\text{model}}$. The following is how these differences are comprehended:
\begin{itemize} 
\item \textbf{$0 \leq \Delta X \leq 2$}:It is impossible to determine whether the model is better because the evidence is \textit{weak}.
\item \textbf{$2 < \Delta X \leq 6$}: The model with the lower value is supported by \textit{positive} evidence. 
\item \textbf{$6 < \Delta X \leq 10$}: The evidence is regarded as \textit{strong}.
\end{itemize}
In order to ensure accuracy and precession in cosmological inference, these statistical diagnostics are crucial requirements for model comparison. using the datasets and methodology as mentioned in this section we showcase the constraints on the model parameters and the cosmological dynamics that put forward under it's light.

\section{Constraints and Cosmological Implications}\label{s6} 
\begin{figure*}[htb]
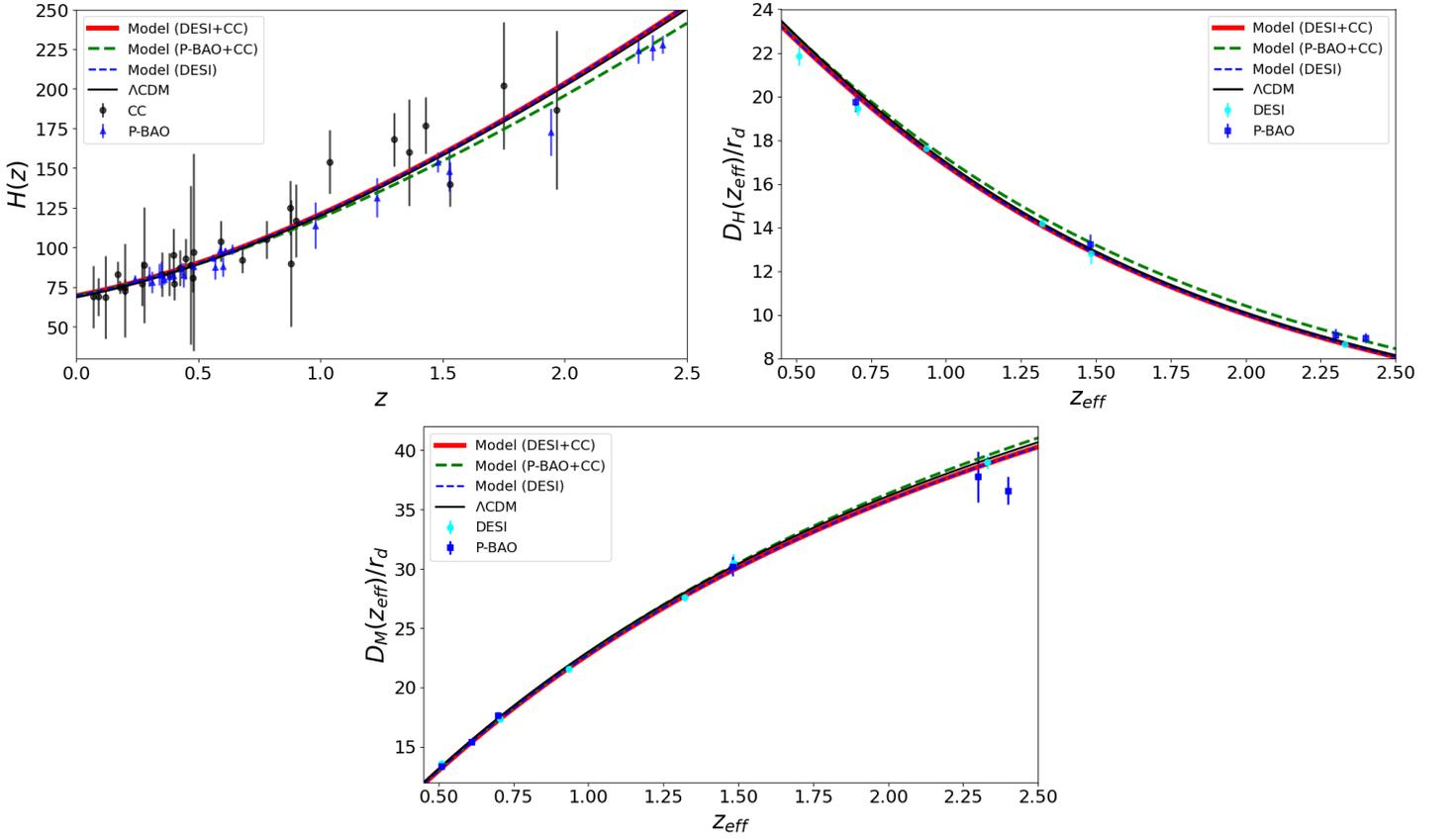

\centerline{
\includegraphics[width=.53\textwidth]{f1a}
\includegraphics[width=.53\textwidth]{f1b}}
\centerline{
\includegraphics[width=.53\textwidth]{f1c}}
\caption{The plot of $H(z)$ vs. $z$, along with $D_M(z_{eff})/r_d$ and $D_H(z_{eff})/r_d$ vs. $z_{eff}$, for the model parameters that best suit the observational data is displayed above. Additionally, a comparison with the $\Lambda$CDM model is shown for each plot.}
\label{f1}
\end{figure*}

\begin{figure*}[htb]
\centerline{
\includegraphics[width=0.7\textwidth]{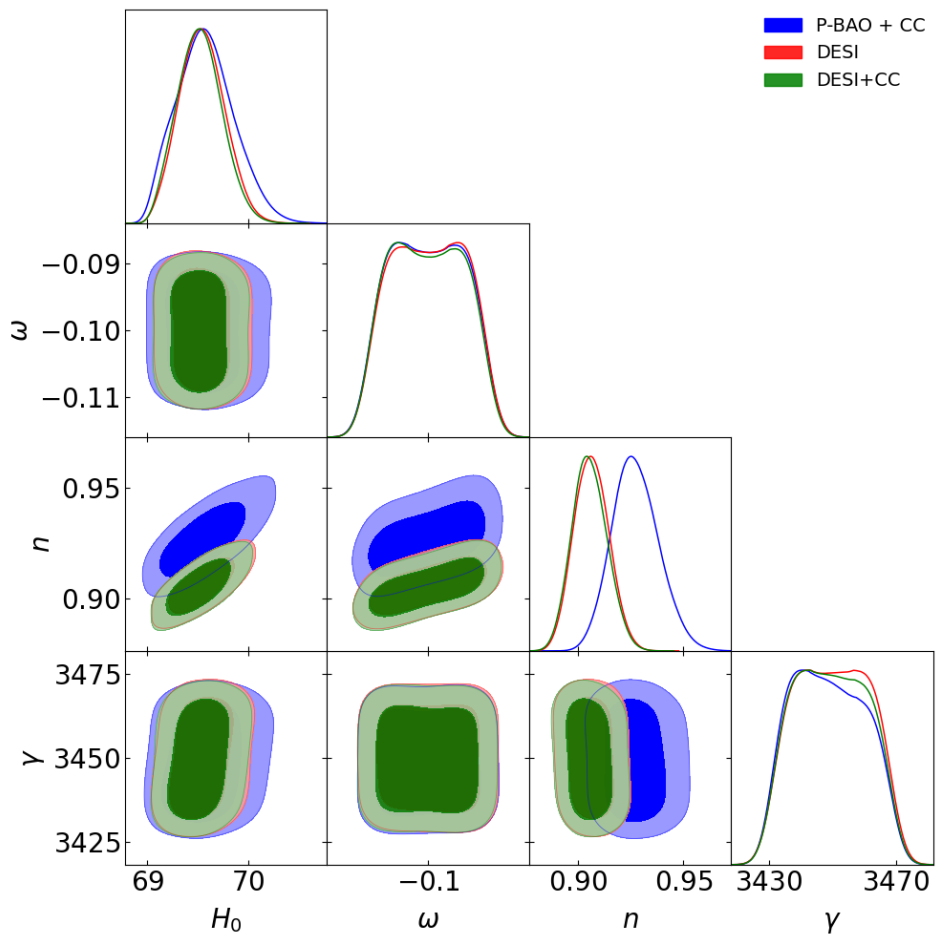}}
\caption{2-d contour sub-plot for the parameters $H_0$, $\omega_{m_0}$, and $\gamma$ with 1-$\sigma$ and 2-$\sigma$ errors (displaying the 68\% and 95\% c.l.) for $H(z)$ vs. $z$.}
\label{f2}
\end{figure*}

Based on the cosmological datasets and the statical analysis approach described in the previous part, this section addresses the parameter constraints and cosmological implications for the corresponding $f(Q,L_m)$ gravity taken into consideration in this study. Here, Tables~(\ref{tab2}) and (\ref{tab3}), along with Figs.~(\ref{f1}) and (\ref{f2}) display the main results. Where, table~(\ref{tab2}) summarises the best-fit values for the model parameters of the aforementioned cosmological model. And, the performance of the given cosmological model against observational data under the light of the constraints on the respective model parameters as mentioned in Table~(\ref{tab2}) as shown in Fig.~(\ref{f1}), which show excellent agreement with observational datasets. The 2D posterior contours for the vital model parameters ($H_0$, $\omega$, $\gamma$ and $n$) with $1\sigma$ and $2\sigma$ confidence regions are shown in Fig.~(\ref{f2}), indicating that the parameters are tightly constrained. Moreover, from the results one can clearly observe that the the constraints become more tighter by the introduction of the DESI dataset as compared to  P-BAO dataset.

\begin{table*}[ht!]
\centering
\renewcommand{\arraystretch}{1.4}
\begin{tabular}{lcccc}
\toprule
\textbf{Dataset} 
& $\boldsymbol{H_0}$ 
& $\boldsymbol{\omega}$ 
& $\boldsymbol{n}$ 
& $\boldsymbol{\gamma}$ \\
\midrule
DESI 
& $69.52^{+0.24}_{-0.23}$ 
& $-0.0963^{+0.0327}_{-0.0362}$ 
& $0.9086^{+0.0285}_{-0.0288}$ 
& $3449.16^{+13.48}_{-13.17}$ \\[4pt]

DESI + CC 
& $69.52^{+0.23}_{-0.23}$ 
& $-0.0975^{+0.0331}_{-0.0352}$ 
& $0.9076^{+0.0276}_{-0.0285}$ 
& $3449.82^{+13.70}_{-13.36}$ \\[4pt]

P-BAO + CC 
& $69.54^{+0.31}_{-0.30}$ 
& $-0.0978^{+0.0328}_{-0.0354}$ 
& $0.9279^{+0.0308}_{-0.0304}$ 
& $3448.69^{+13.60}_{-12.88}$ \\
\bottomrule
\end{tabular}
\caption{Model parameter's best-fit values with $1\sigma$ confidence intervals were derived from the DESI, DESI + CC, and P-BAO + CC datasets.}
\label{tab2}
\end{table*}

\begin{table*}[ht!]
\centering
\renewcommand{\arraystretch}{1.25}
\begin{tabular}{llcccccc}
\toprule
\textbf{Model} & \textbf{Dataset} 
& $\boldsymbol{\chi^2_{\min}}$ 
& \textbf{AIC} 
& \textbf{BIC} 
& $\boldsymbol{R^2}$ 
& $\boldsymbol{\Delta AIC}$ 
& $\boldsymbol{\Delta BIC}$ \\
\midrule
{\textbf{$f(Q,L_m)$ Model}} 
 & DESI     & 10.40 & 18.40 & 20.34 & 0.9980 & 8.82 & 7.84 \\
 & DESI + CC     & 25.14 & 33.14 & 40.18 & 0.9980 & 8.98 & 5.47 \\
 & P-BAO + CC    & 42.63 & 50.63 & 59.45 & 0.9895 & 3.84 & $-0.57$ \\
\midrule
{\boldmath$\Lambda$CDM} 
 & DESI    & 23.22 & 27.22 & 28.18 & 0.9968 & -- & -- \\
 & DESI + CC     & 38.12 & 42.12 & 45.65 & 0.9968 & -- & -- \\
 & P-BAO + CC    & 50.47 & 54.47 & 58.88 & 0.9787 & -- & -- \\
\bottomrule
\end{tabular}
\caption{Statistical comparison between the $f(Q,L_m)$  gravity model and the standard $\Lambda$CDM cosmology using DESI, DESI + CC, and P-BAO + CC datasets. The quantities $\Delta\mathrm{AIC}$ and $\Delta\mathrm{BIC}$ are defined as $\mathrm{AIC}_{\Lambda\mathrm{CDM}}-\mathrm{AIC}_{\text{model}}$ and $\mathrm{BIC}_{\Lambda\mathrm{CDM}}-\mathrm{BIC}_{\text{model}}$, respectively.}
\label{tab3}
\end{table*}
\begin{table}[h!]
\centering
\renewcommand{\arraystretch}{1.25}
\begin{tabular}{lccc}
\toprule
Dataset 
& $q(0)$ 
& $w_{\mathrm{eff}}(0)$ 
& $z_{\mathrm{tr}}$ \\
\midrule
DESI 
& $-0.5683^{+0.0029}_{-0.0030}$ 
& $-0.7122^{+0.0019}_{-0.0020}$ 
& $0.6957^{+0.0039}_{-0.0024}$ \\

DESI + CC 
& $-0.5687^{+0.0030}_{-0.0026}$ 
& $-0.7125^{+0.0020}_{-0.0017}$ 
& $0.6962^{+0.0023}_{-0.0042}$ \\

P--BAO + CC 
& $-0.5790^{+0.0048}_{-0.0051}$ 
& $-0.7194^{+0.0032}_{-0.0034}$ 
& $0.7673^{+0.0066}_{-0.0049}$ \\
\bottomrule
\end{tabular}
\caption{Values at $z=0$ with $1\sigma$ uncertainties and the transition redshift $z_{\mathrm{tr}}$ obtained from different observational datasets.}
\label{tab4}
\end{table}
Table~(\ref{tab3}) presents a detailed statistical comparison between the given $f(Q,L_m)$ gravity model and the standard $\Lambda$CDM cosmology using DESI, DESI + CC, and P-BAO + CC datasets. From the table, it is evident that the $f(Q,L_m)$ model consistently yields lower values of the minimum chi-square $\chi^2_{\min}$ along with higher coefficients of determination $R^2$ compared to $\Lambda$CDM, demonstrating an improved overall fit to the observed data. In addition, the corresponding AIC and BIC values are systematically smaller for the $f(Q,L_m)$ model, reflecting its improved explanatory power even after accounting for the number of free parameters. The statistical preference for the $f(Q,L_m)$ model can be further quantified through the information criteria differences, given by $\Delta\mathrm{AIC}$ and $\Delta\mathrm{BIC}$. For the DESI dataset, the obtained values $\Delta$AIC = 8.82 and $\Delta$BIC = 7.84 provide strong evidence in favour of the $f(Q,L_m)$ model over the standard $\Lambda$CDM scenario. A similar trend is observed for the DESI + CC dataset, where $\Delta$AIC = 8.98 and $\Delta$BIC = 5.47 again indicate substantial statistical support for the modified gravity framework. For the P-BAO + CC dataset, the positive value $\Delta$AIC = 3.84 suggests moderate evidence in favour of the $f(Q,L_m)$ model, whereas the slightly negative value $\Delta$BIC = -0.57 implies that, once the stronger BIC penalty is applied, both models perform at a nearly comparable level as often BIC is seen to put more severe penalty. Nevertheless, when all datasets are considered collectively, the overall trend clearly favours the $f(Q,L_m)$ model as a statistically competitive and, in most cases, superior alternative to $\Lambda$CDM, providing an improved description of late-time cosmological observations without the explicit introduction of a cosmological constant.\\
Finally, under the light of the constraints on the model parameter we analyse the evolution of the deceleration parameter, effective EoS, and statefinder diagnostics for the model using Eqs.~(\ref{eq1}), (\ref{ew1}), (\ref{es}), and (\ref{er}) as shown in the following part of this section. In addition, we assess the corresponding energy conditions for the give gravity model in order to demonstrate how well it complies with observational constraints. To further investigate the model's viability in terms of dynamical stability, we conduct a stability study of the model under homogeneous linear perturbations.

\subsection{Deceleration Parameter}
\begin{figure*}[htb]
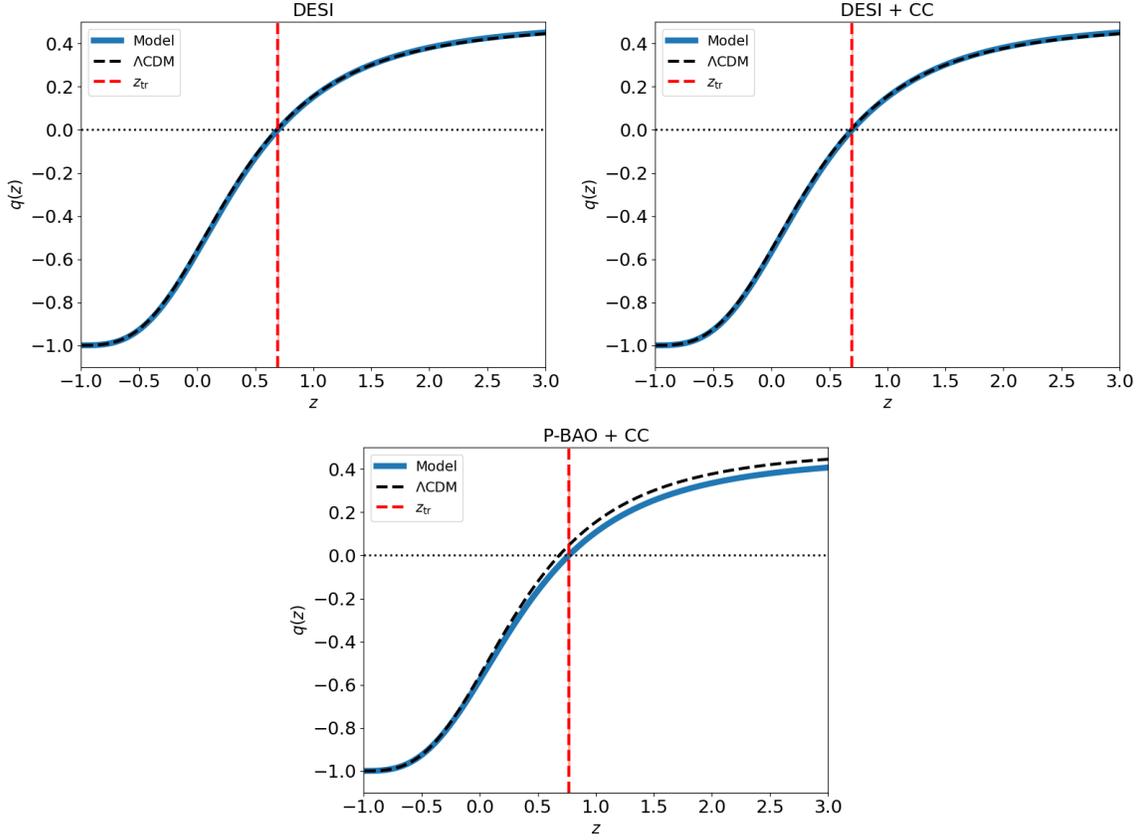

\centerline{
\includegraphics[width=.42\textwidth]{f3a}
\includegraphics[width=.42\textwidth]{f3b}}
\centerline{
\includegraphics[width=.42\textwidth]{f3c}
}
\caption{Plot showing evolution of Deceleration parameter with redshift and the corresponding transitional redshift for best-fit parameter values of the model.}
\label{f3}
\end{figure*}
One important paramter that depicts the nature of the expansion history of the Universe is the deceleration parameter $q(z)$, which indicates whether the cosmic expansion is accelerating or decelerating at a certain redshift. In contrast to a positive value of $q(z)$, which is expected during the radiation- and matter-dominated epochs when gravity dominates and corresponds to a decelerated expansion, a negative value of $q(z)$ indicates an accelerated expansion due to a repulsive component like dark energy or an effective modification of gravity. Here, the epoch at which the Universe undergoes a transition from deceleration to acceleration thats a feature of late-time cosmic history, is indicated by the transition redshift $z_{\rm tr}$, which is defined as the redshift for which $q(z) = 0$. Figure~\ref{f3} shows the redshift evolution of the deceleration parameter $q(z)$ for the given model constrained using different datasets. At higher redshifts, all trajectories for $q(z)$ approach a positive values, indicating a decelerated expansion phase thats consistent with a matter-dominated Universe. But, when the redshift decreases, the evolution of $q(z)$ shows a change in behaviour, marking a transition from deceleration to acceleration phase, as it changes from a positive to negative value. This transition is seen to occurs at a redshift \(z_{\rm tr} \simeq 0.69\text{--}0.76\), with slight variations depending on the dataset, showcasing the sensitivity of the transition epoch to the underlying observational constraints. However, at the present epoch ($z = 0$), the deceleration parameter is seen to be negative for all dataset, providing robust evidence for the current accelerated expansion of the Universe. Results indicates $q(0) \in [-0.5790, -0.5683]$, which agree with current research on the Hubble tension \cite{verde2019tensions} and independent observational estimates from Planck \cite{planck2018}. Here, we observe the P-BAO+CC dataset favors a slightly stronger late-time acceleration, while the DESI and DESI+CC cases exhibit a nearly milder present-day accelerated expansion. Also, the evolution of \(q(z)\) constrained by the DESI and DESI+CC datasets exhibits a quite closer resemblance to the \(\Lambda\)CDM behaviour than that inferred from the P-BAO+CC dataset. All curves convergence towards $q \to -1$ for $z \to -1$, indicating a de Sitter-like future evolution. For all datasets, the transition redshift $z_{\text{tr}}$ are tightly constrained placing the very beginning of cosmic acceleration firmly within the range $z_{\rm tr} \in [0.6957, 0.7673]$, which is in good agreement with previous and recent studies that estimate the beginning of acceleration in vicinity of $z \sim 0.6 - 0.8$ \cite{farooq2013, xu2012, xu2012new}. Overall, evolution of $q(z)$ demonstrates that the $f(Q,L_m)$ gravity model naturally reproduces the key features of late-time cosmic expansion, which includes a matter-dominated decelerating phase at high redshift, a smooth and observationally consistent transition to acceleration at intermediate redshifts, and a sustained accelerated expansion at the present epoch without invoking an explicit cosmological constant.

\subsection{Effective EoS}
\begin{figure*}[htb]
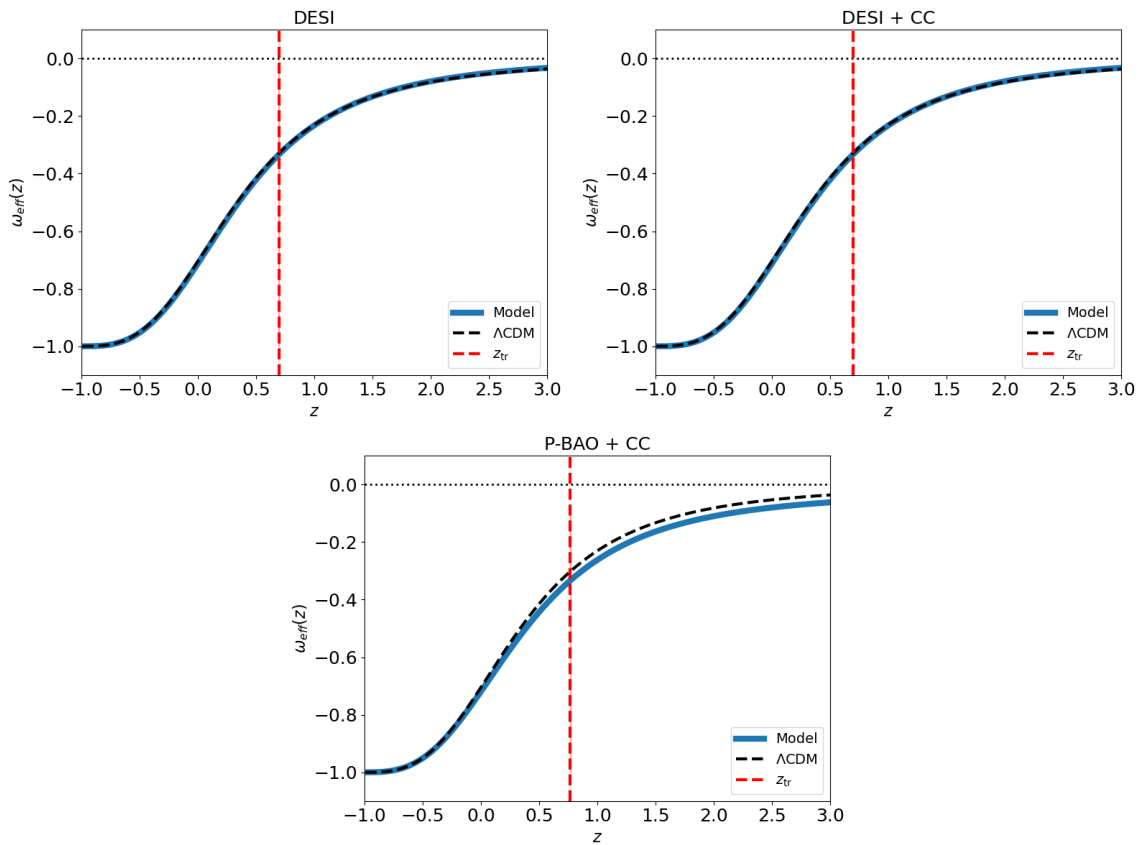

\centerline{
\includegraphics[width=.42\textwidth]{f4a}
\includegraphics[width=.42\textwidth]{f4b}}
\centerline{
\includegraphics[width=.42\textwidth]{f4c}
}
\caption{Plot showing the redshift evolution of effective EoS for best-fit parameter values for the given model.}
\label{f4}
\end{figure*}

The dynamical evolution of the cosmic fluid can be effectively characterized by the effective equation-of-state (EoS) parameter $\omega_{\mathrm{eff}}(z)$, which captures the combined effects of radiation, matter, and geometrical effects, arising within the $f(Q,L_m)$ gravity framework. When $\omega_{\mathrm{eff}}(z) < -\frac{1}{3}$, accelerated cosmic expansion occurs, whereas the ordinary $\Lambda$CDM scenario is represented by the limiting case $\omega_{\mathrm{eff}}(z) = -1$. A phantom regime is indicated by values of $\omega_{\mathrm{eff}}(z) < -1$, whereas a quintessence-like phase is characterised by $-1 < \omega_{\mathrm{eff}}(z) \leq -\frac{1}{3}$, and a Chaplygin-like phase is characterised by $-1 \leq \omega_{\mathrm{eff}}(z) \leq 0$. The evolution of $\omega_{\mathrm{eff}}(z)$ clearly shows a smooth transition from a matter-dominated decelerated phase at high redshift to a late-time accelerated expansion, as seen in Fig.~(\ref{f4}). One can observe that $\omega_{\mathrm{eff}}$ is an increasing function of the redshift, and it tends to the $\Lambda$CDM value when $z \to -1$. At early epochs, $\omega_{\rm eff}(z)$ approaches values close to zero, consistent with a pressureless matter-dominated Universe, while at low redshifts it crosses below $-1/3$, thereby triggering cosmic acceleration. For the current era (\(z=0\)), all dataset combinations yield values of the effective EoS well below the acceleration threshold \(\omega_{eff} = -1/3\), namely in the range $\omega_{eff} \sim -0.71$, consistent with current dynamical dark energy scenarios \cite{verde2019tensions} and in good agreement with observational constraints from Planck 2018 \cite{planck2018} as well. This shows that the Universe is currently undergoing an accelerated expansion of quintessence or Chaplygin type. Among the datasets, the P-BAO+CC combination favors a marginally stronger late-time acceleration, reflected in its slightly more negative value of \(\omega_{\rm eff}(0)\), whereas the DESI and DESI+P-BAO based constraints indicate a comparatively milder acceleration. In particular, \(\omega_{eff}(z)\) constrained by the DESI and DESI+CC datasets exhibits a closer resemblance to the \(\Lambda\)CDM behavior than that inferred from the P-BAO+CC dataset. Specifically, similar to $q(z)$, $\omega_{eff}(z)$ supports a sustained accelerated expansion at the current epoch without requiring an explicit cosmological constant, as well as a smooth and observationally compatible transition to acceleration at intermediate redshifts.
\subsection{Statefinder Diagnostics}
\begin{figure*}[htb]
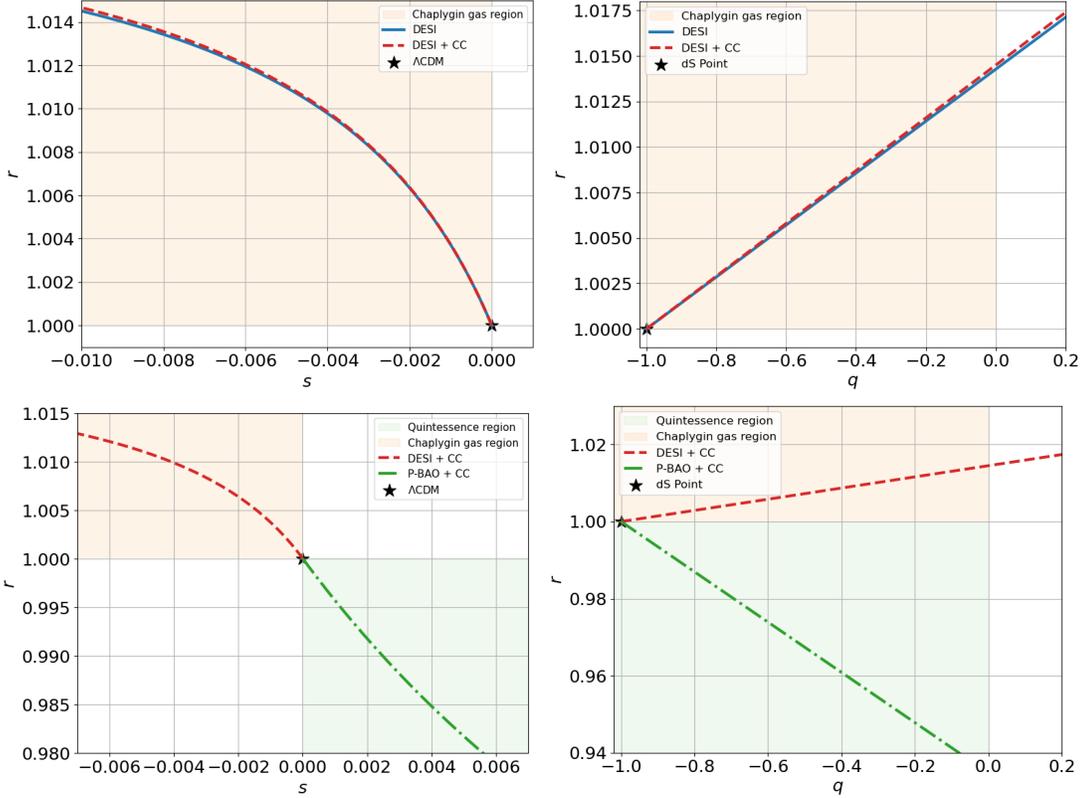

\centerline{
\includegraphics[width=.4\textwidth]{f5a}
\includegraphics[width=.4\textwidth]{f5b}}
\centerline{
\includegraphics[width=.4\textwidth]{f5c}
\includegraphics[width=.4\textwidth]{f5d}
}
\caption{Plot showing the evolution of the statefinder diagnostic pairs with redshift using best-fit parameter values for the given model.}
\label{f5}
\end{figure*}

Since the statefinder pairs $(r,s)$ and $(r,q)$ have a geometrical nature, the statefinder parameters serve as universal diagnostics for analyzing dark energy models \cite{ref97n}. We analyze the evolutionary trajectories of the cosmological model in the $(r,s)$ and $(r,q)$ planes using the constraints obtained from the observational datasets. The fixed point $(r,s)=(1,0)$ corresponds to the $\Lambda$CDM model, while the de~Sitter (dS) phase is represented by the point $(r,q)=(1,-1)$. Figure~\ref{f5} shows that, in the $(r,s)$ plane, the model approaches the $\Lambda$CDM fixed point, $(r,s)=(1,0)$, at late times for all dataset combinations. Similarly, in the $(r,q)$ plane, the model consistently evolves toward the de~Sitter point $(r,q)=(1,-1)$ in the late-time regime. The recent cosmic phase transition is further supported by the signature flip of the deceleration parameter from some positive to a negative values in the plots. However, evolution of both the diagnostics pair shows that for constraints from DESI and DESI+CC the model approaches the $\Lambda$CDM model and de~Sitter phase in a Chaplygin gas like manner, whereas for P-BAO +CC the model approaches the $\Lambda$CDM model and de~Sitter phase through a quintessence dominated evolution. 

\subsection{Energy Conditions}
Here, we further evaluate the physical feasibility of the $f(Q)$ gravity model by analysing the behaviour of the classical energy requirements as functions of the redshift $z$. In order to make certain that the effective energy density and pressure resulting from the modified gravitational field equations stay within physically plausible limitations, these energy requirements are crucial consistency criteria. The conditions are formulated in terms of the effective energy density $\rho_{\mathrm{eff}}$ and effective pressure $p_{\mathrm{eff}}$ as follows:
\begin{itemize}
    \item \textbf{Weak Energy Condition (WEC)}: $\rho_{\mathrm{eff}} \geq 0$, \hspace{3mm} $\rho_{\mathrm{eff}} + p_{\mathrm{eff}} \geq 0$,
    \item \textbf{Null Energy Condition (NEC)}: \hspace{6mm} $\rho_{\mathrm{eff}} + p_{\mathrm{eff}} \geq 0$,
    \item \textbf{Strong Energy Condition (SEC)}: \hspace{3mm} $\rho_{\mathrm{eff}} + 3p_{\mathrm{eff}} \geq 0$,
    \item \textbf{Dominant Energy Condition (DEC)}: \hspace{2mm} $\rho_{\mathrm{eff}} - p_{\mathrm{eff}} \geq 0$.
\end{itemize}
In this context, the modified Friedmann equations related to the $f(Q)$ cosmological model yield the effective energy density and pressure, which are given by:
\begin{equation}
3H^{2} = \rho_{\mathrm{eff}},
\label{q1}
\end{equation}
\begin{equation}
2\dot{H} + 3H^{2} = -p_{\mathrm{eff}},
\label{q2}
\end{equation}
where $H$ denotes the Hubble parameter. These relations enable a systematic investigation of the redshift evolution of the effective cosmic fluid induced by the reconstructed geometry and allow us to determine whether the classical energy conditions are satisfied or violated throughout the cosmic history.

\begin{figure*}[htb]
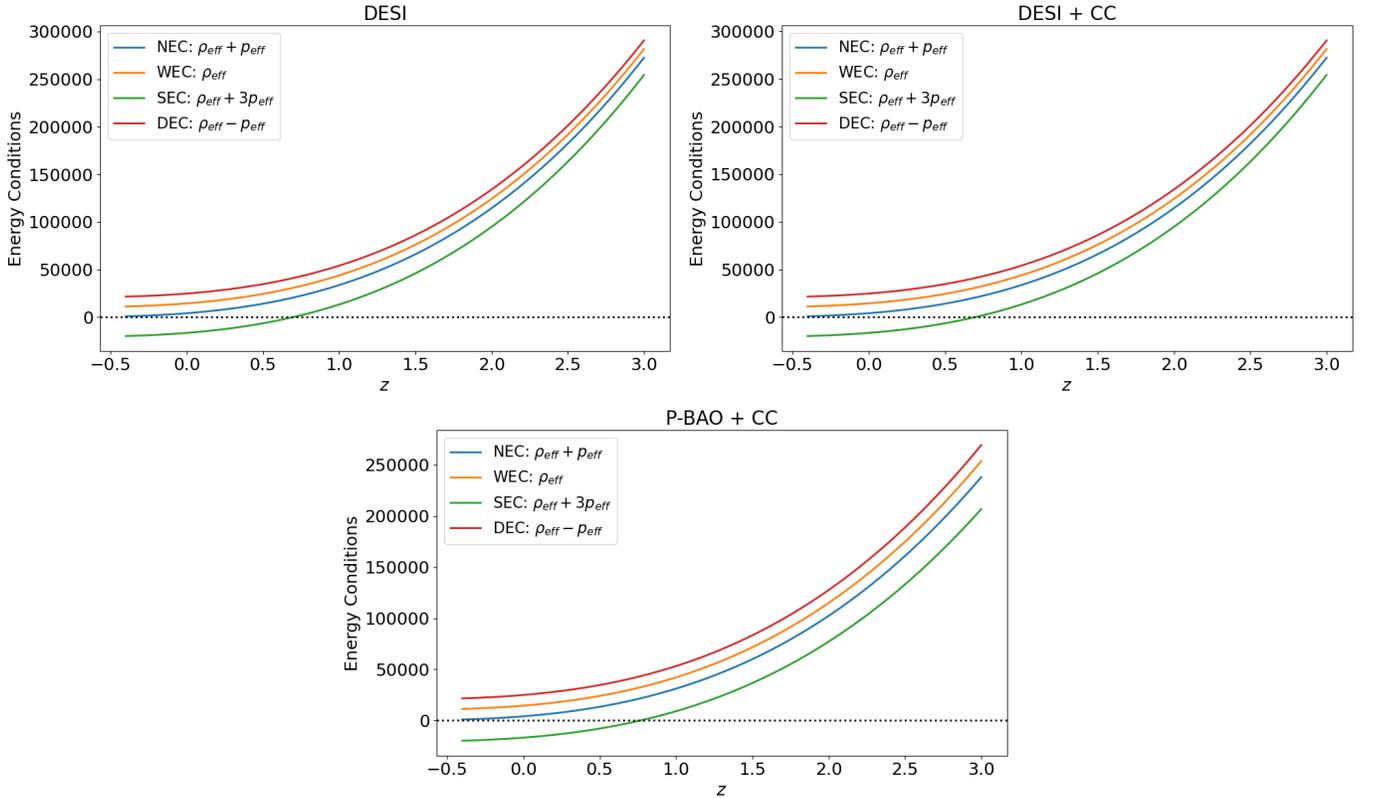

\centerline{
\includegraphics[width=.5\textwidth]{f6a}
\includegraphics[width=.5\textwidth]{f6b}}
\centerline{
\includegraphics[width=.5\textwidth]{f6c}
}
\caption{Plot showing redshift evolution of the energy conditions for best-fit parameter values for the given model.}
\label{fe}
\end{figure*}
Using the model parameters' best-fit values we investigate the redshift evolution of the classical energy conditions by employing Eq.~(\ref{Hz}) in conjunction with Eqs.~(\ref{q1}) and (\ref{q2}). The corresponding behavior of the energy conditions is illustrated in Fig.~\ref{fe}. It is evident that both the Weak Energy Condition (WEC) and the Dominant Energy Condition (DEC) are fulfilled for the whole course of the cosmic development. This indicates a physically permissible and non-exotic effective cosmic fluid by ensuring that the effective energy density is positive and that the effective pressure does not surpass the energy density. Moreover, the model preserves the Null Energy Condition (NEC) over the full redshift range. This shows that the effective cosmic fluid remains physically admissible and absolutely free from phantom instabilities, while supporting a smooth and dynamically stable late-time cosmic acceleration. On the other hand, at some lower redshifts ($z \lesssim z_{tr}$), the Strong Energy Condition (SEC) is seen to be broken, which is consistent with the start of the present accelerating expansion of the Universe. However, the SEC become valid at higher redshifts ($z \gg z_{tr}$), suggesting a return to a matter-dominated, decelerating epoch. This results all together highlights the model’s ability to explain cosmic acceleration as a purely geometric effect while remaining physically consistent and in good agreement with observational constraints.

\subsection{Stability Analysis}
The dynamic stability of the proposed $f(Q,\mathcal{L}_m)$ gravity model against homogeneous linear perturbations will be investigated in this section. To this end, we introduce first-order perturbations in the Hubble parameter and the effective energy density \cite{ref90AN,ref90AN1}, such that:
\begin{equation}
H^{*}(t) = H(t)\,[1 + \delta_1(t)],
\label{An1}
\end{equation}
\begin{equation}
\rho_{eff}^{*}(t) = \rho_{eff}(t)\,[1 + \delta_2(t)],
\label{An2}
\end{equation}
where $H^{*}(t)$ and $\rho_{eff}^{*}(t)$ denote the Hubble parameter and effective energy density that are perturbed, while $\delta_1(t)$ and $\delta_2(t)$ represent their respective perturbations. In standard cosmology, the conservation of the effective energy--momentum tensor is governed by the continuity equation
\begin{equation}
\dot{\rho_{eff}} + 3H(\rho_{eff} + p_{eff}) = 0,
\label{An3}
\end{equation}
where differentiation with respect to cosmic time $t$ is shown by the overdot. The time derivative can be conveniently expressed in terms of the redshift $z$ through the relation:
\begin{equation}
\frac{d}{dt} = - (1 + z)\, H(z)\, \frac{d}{dz}.
\label{r1}
\end{equation}
\begin{figure*}[htb]
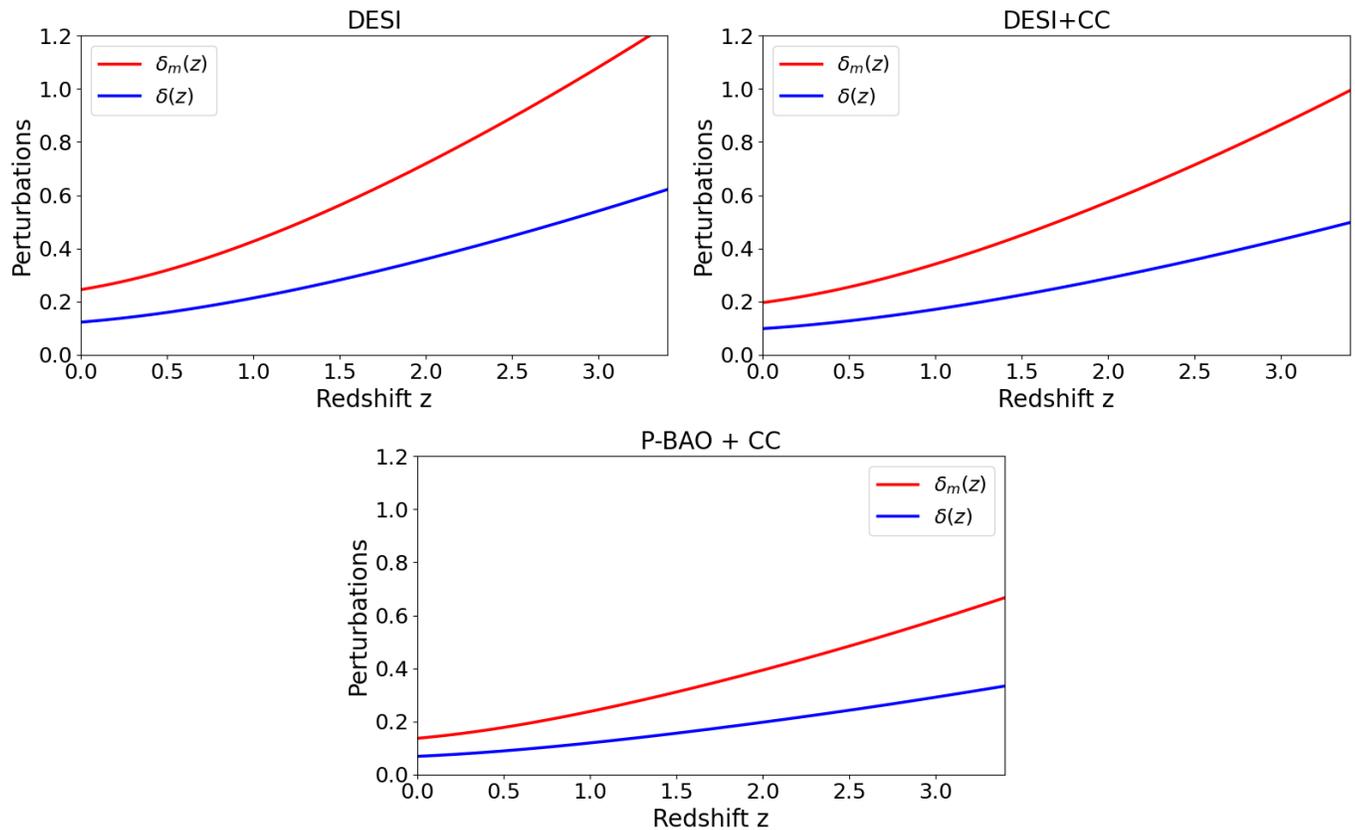

\centerline{
\includegraphics[width=.5\textwidth]{f7a}
\includegraphics[width=.5\textwidth]{f7b}}
\centerline{
\includegraphics[width=.5\textwidth]{f7c}
}
\caption{Plot showing the evolution of the perturbation terms $\delta_1(z)$ and $\delta_2(z)$ for each dataset's best model parameters as a function of redshift $z$.}
\label{f7}
\end{figure*}
Substituting Eqs.~(\ref{An1}), (\ref{An2}), (\ref{ew1}), and (\ref{q1}) into Eq.~(\ref{An3}), and retaining only linear terms in the perturbations, we obtain the following first-order relations:
\begin{equation}
\dot{\delta}_2(t) + 3H\,(1 + \omega_{\rm eff})\, \delta_1(t) = 0,
\label{An4}
\end{equation}
\begin{equation}
2\,\delta_1(t) = \delta_2(t).
\label{An5}
\end{equation}
Using Eq.~(\ref{r1}) and transforming the above equations into redshift space, we arrive at
\begin{equation}
-(1 + z)\, \frac{d\delta_2}{dz} + \frac{2}{3}\left(1 + \omega_{\rm eff}(z)\right)\delta_2 = 0,
\label{An4a}
\end{equation}
\begin{equation}
2\,\delta_1(z) = \delta_2(z),
\label{An5a}
\end{equation}
where the effective equation-of-state parameter is given by:
\begin{equation}
\omega_{\rm eff}(z) = -1 + \frac{2(1 + z)}{3H(z)} \frac{dH}{dz}.
\end{equation}
Solving Eq.~(\ref{An4a}), we trace the redshift evolution of the perturbation variables $\delta_1(z)$ and $\delta_2(z)$. Using the best-fit parameter values of the $f(Q,L_m)$ gravity model constrained by different observational datasets, the resulting behavior of these perturbations is presented in Fig.~\ref{f7}. The redshift evolution of the perturbation functions $\delta_1(z)$ and $\delta_2(z)$ exhibits a clear monotonic decrease as the redshift $z$ decreases. This behavior indicates that the perturbations gradually decay out as the Universe progress toward the current era. Such smooth damping of the magnitudes of the perturbation implies the absence of any kind of late-time growth of instabilities, confirming that the given $f(Q,L_m)$ gravity model is dynamically stable against such linear homogeneous perturbations. Additionally, the uniform decay pattern found in all observational datasets shows that the model's stability is reliable and unaffected by the choice of dataset. The $f(Q,L_m)$ framework's internal coherence and physical viability are strongly supported by this uniform damping behaviour, which further supports the framework's acceptability as a geometrically driven alternative for explaining the observed late-time accelerated expansion of the Universe.
\section{Conclusion}\label{s7}
In this study, we have used recent observational datasets such as DESI BAO, prior BAO (P-BAO), and cosmic chronometer (CC) datasets to conduct a thorough cosmological study of a well-defined $f(Q,L_m)$ gravity model. We achieved well constrained best-fit values of the model parameters via a Markov Chain Monte Carlo (MCMC) approach and thoroughly examined the resulting background cosmological dynamics, energy conditions, and stability aspects.\\
As Table~(\ref{tab2}) summarizes, the given model shows a highly consistent determination of the Hubble constant across all dataset combinations, with the DESI, DESI+CC, and P-BAO+CC datasets converging to $
H_0 \simeq 69.5~\mathrm{km\,s^{-1}\,Mpc^{-1}}
$ within tight uncertainties. This robustness against the inclusion of cosmic chronometer and BAO data indicates that the inferred expansion rate is largely insensitive to the choice of late-time probes. The obtained value of $H_0$ lies between the early-Universe estimate from Planck \cite{planck2018} and the late-Universe SH0ES determination \cite{Riess2021}, yielding an intermediate expansion rate that may help ease the Hubble tension without invoking exotic physics. Moreover, the results for $H_0$ shows close agreement with recent $f(Q)$ and $f(Q,L_m)$ gravity studies. For example, by utilising a model independent reconstruction of $f(Q)$ gravity Capozziello and D’Agostino~\cite{capozziello2022} obtained $
H_0 \simeq 69.3~\mathrm{km\,s^{-1}\,Mpc^{-1}}$. In the same way, Sakr and Schey~\cite{Sakr} investigated the Hubble tension in the context of the $f(Q)$ framework and found feasible parameterizations that preferred values between $H_0 \simeq 68\text{--}70~\mathrm{km\,s^{-1}\,Mpc^{-1}}$. Study by Y. Myrzakulov et al. \cite{n1} with a $f(Q,L_m)$ gravity model similar to ours proposed $H_0 \simeq 69.96~\mathrm{km\,s^{-1}\,Mpc^{-1}}$. Likewise, we also discovered that the parameter $n$ deviates from unity across all datasets, indicating departures from the standard matter evolution predicted by General Relativity, with the effect being particularly pronounced for DESI and DESI+CC datasets. Also, the relative stability of the parameter $\gamma = \frac{\lambda}{6\alpha} \sim 3449$ across different observational datasets the robustness of the model. The deceleration parameter remains negative at the present epoch for all dataset combinations, while the transition redshift \(z_{\rm tr}\) lies in the range \(z_{\rm tr} \simeq 0.69\text{--}0.76\), which is fully consistent with recent observational determinations and closely aligned with the predictions of the standard \(\Lambda\)CDM model \cite{pp47,n1, capozziello2022,Sakr}. Notably, the inclusion of DESI data leads to tighter constraints and slightly milder acceleration, clearly illustrating the impact of high-precision BAO measurements on late-time cosmic dynamics for the given model. The redshift evolution of the effective equation-of-state \(\omega_{\rm eff}(z)\) shows a smooth and physically consistent transition from a matter-dominated decelerated era at high redshifts to an accelerated phase at lower redshifts or late times. Further, the statefinder diagnostics showcases that the model for both DESI and DESI+CC datasets indicate a Chaplygin gas like behaviour, whereas the P-BAO+CC dataset favours a quintessence-like behavior in it's late time evolution. However, for all the datasets the model's trajectories finally approaches the $\Lambda$CDM model and de~Sitter phase in the far future. Also, the detailed analysis of the classical energy conditions validates the physical consistency of the model. The Weak Energy Condition (WEC) are satisfied at all cosmic epochs just like the Dominant Energy Condition (DEC), that ensures a positive effective energy density and a well-behaved effective cosmic fluid, while the Null Energy Condition (NEC) remains preserved throughout cosmic evolution. However, the Strong Energy Condition (SEC) is seen to violate at low redshifts, as required to account for the universe's observed accelerated expansion, and is restored at higher redshifts, consistently recovering the standard matter-dominated decelerating phase. For stability analysis the perturbation variables \(\delta_1(z)\) and \(\delta_2(z)\) depicts a monotonic decay toward lower redshifts for all observational datasets considered, indicating the absence of growing modes or instabilities and confirming the stability of the background cosmological solution. The robustness of this damping behavior across different cosmological datasets underscores the internal consistency of the \(f(Q,L_m)\) model and aligns with similar stability analyses frequently reported in literatures on non-metricity-based modified gravity models.\\
From a broader point of view, our results support the notion that the late-time evolution of the universe can be significantly influenced by matter--geometry coupling. We demonstrate that the late time cosmic acceleration may be viewed as a purely geometric event arising from non-metricity effects instead of a \emph{ad hoc} cosmological constant or exotic dark energy fluid, which is consistent with a number of recent investigations. By demonstrating that such models remain observationally feasible, energetically consistent, and dynamically stable when confronted with the most recent high-precision datasets, such as DESI DR2, our work attempted to expand on the previous findings. Further studies incorporating structure growth, redshift-space distortions, and upcoming data surveys such as \textit{Euclid}\cite{Euclid2018}, \textit{LSST}\cite{LSST2019}, and \textit{SKA} \cite{SKA2015} will be crucial for testing such frameworks and distinguishing it from standard cosmology, potentially which can offer deeper insight into the geometric origin of cosmic acceleration.

\end{document}